%% file: main_nonanon.tex
\documentclass[twocolumn]{aastex631}

\usepackage[normalem]{ulem}
\usepackage{soul}

\accepted{to PASP}

\shortauthors{Salazar et al.}
\graphicspath{{./}{figures/}}

\begin{document}

\title{Automated Detection of Double-Lined Spectroscopic Binaries in High-Resolution Spectra and a Probabilistic Analysis of Stellar Multiplicity}

\correspondingauthor{Steven Giacalone}
\email{giacalone@astro.caltech.edu}

\author{William B. Salazar}
\affiliation{Department of Physics and Astronomy, California State Polytechnic University Pomona, 3801 West Temple Avenue, Pomona, CA 91768, USA}

\author[0000-0002-8965-3969]{Steven Giacalone}
\altaffiliation{NSF Astronomy and Astrophysics Postdoctoral Fellow}
\affiliation{Department of Astronomy, California Institute of Technology, Pasadena, CA 91125, USA}

\author[0000-0001-8638-0320]{Andrew W. Howard}
\affiliation{Department of Astronomy, California Institute of Technology, Pasadena, CA 91125, USA}



\begin{abstract}

The characterization of Sun-like stars, including determining whether those stars are single or have stellar companions, has become increasingly important for missions seeking planets beyond the Solar System. With upcoming missions like the Habitable Worlds Observatory (HWO) on the horizon, vetting stars for binarity and confirming potential targets as viable is essential precursor science. In this paper, we present an automated algorithm that searches for double-lined spectroscopic binaries (SB2s) in high-resolution spectra. Using MOLUSC, we simulate realistic stellar populations to quantify the capability of algorithm for detecting stellar companions around nearby stars, with potential applications to HWO target stars and Gaia DR4 planet-host candidates. In addition, we explore how the detection rate is influenced by complementary observations like high-resolution adaptive optics imaging and time-series radial velocities. We find that our SB2 detection algorithm alone can rule out $46.1\%$ of stellar-mass companions. High-resolution adaptive optics alone rules out $79.6\%$ of stellar companions. Radial velocity surveys alone ruled out $98.4\%$ and $98.7\%$ for 5-year and 10-year surveys, respectively. When all three methods are combined, the probability of a typical nearby star having a stellar companion is $<1\%$. The combination of the methods allows us to better understand and resolve target lists for future surveys. 

\end{abstract}


\section{Introduction} \label{sec:intro}

Binary stars have a profound impact on our understanding of the universe; the presence of unseen stellar companions can greatly impact the estimated parameters of the primary stars based on spectroscopy \citep{elbadry2018implications, furlan2020unresolved} and therefore the parameters and statistics of orbiting planets \citep{ciardi2015multiplicity, hirsch2017companions, savel2020multiplicity, moe2017pq, sullivan2022bbinary, sullivan2022abinary, sullivan2023binary, sullivan2024binary, sullivan2026binary, bergsten2026}. When combined with the fact that roughly half of all Sun-like stars have at least one stellar companion \citep{raghavan2010multiplicity, moe2017pq, gonzalezpayo2026}, identifying these multi-star systems becomes imperative.

Binary stars than cannot be resolved spatially but can be resolved spectroscopically have historically been divided into two categories: single-lined spectroscopic binaries (SB1), for which the lines of only one star are resolved in the spectrum and and the presence of an additional star is inferred based on shifts in radial velocity, and double-lined spectroscopic binaries (SB2), for which the lines of both stars are resolved in the spectrum. The development of techniques for detecting and characterizing SB2s spans decades, with the earliest works focusing on the computation of a cross-correlation function (CCF) between the spectrum and a stellar line mask \citep{simkin1974ccf, tonry1979ccf}. With this technique, SB2s are identifiable as a double-peaked CCF. Later, techniques for determining the properties and orbits of the two stellar components in double-lined spectra were introduced by works like \citet{zucker1994todcor} and \citet{simon1994disentangling}.

As time progressed and the availability of spectroscopic data increased, new and more sophisticated tools emerged for identifying SB2s. Within the last decade, surveys like LAMOST \citep{luo2015LAMOST}, SDSS \citep{york2000SDSS}, and DESI \citep{desi2024DESI} have swept nearly the entire sky, obtaining low-to-medium-resolution spectroscopy of millions of stars. To identify SB2s in these large data sets, many have turned to autonomous disentangling algorithms \citep{seeburger2024disentagling, gonzalez2024disentangling} and data-driven models \citep{elbadry2018discovery, li2025mining, wang2025deeplearning} with high success.

However, large training sets are not always available for detecting SB2s. For example, targeted ground-based surveys of stars with high-resolution ($R \sim 100,000$) spectrographs must often search for these binaries in a non-data-driven way. In this situation, the CCF technique continues to be the best option for identifying SB2s. For instance, \cite{kolbl2015reamatch} developed a technique for efficiently searching for SB2s in Keck/HIRES data using a CCF-based technique. However, this method relies on manual inspection, which complicates the quantification of detection completeness and false alarm rates. 

The need for systematic binary star characterization has become particularly acute in the context of next-generation direct imaging missions. For instance, the planned Habitable Worlds Observatory (HWO; \citealt{NAS2023hwo}) aims to directly image and characterize Earth-like planets around nearby Sun-like stars using high-contrast imaging techniques. The presence of undetected stellar companions can significantly complicate or even preclude the detection of habitable zone planets, making thorough vetting of potential target stars essential \citep{mamajek2023hwo, harada2024hwo, tuchow2025hwo, fetherolf2026}. However, many stars on preliminary HWO target lists have not been comprehensively screened for stellar companions, and their brightness often exceeds the {\it Gaia} saturation limit \citep{gaia2016}, necessitating ground-based follow-up observations.

In this work, we present an automated algorithm for detecting SB2s in high-resolution spectra with well-characterized detection completeness and false alarm rates. Building on the traditional CCF approach, we develop a framework that enables statistically rigorous binary star detection through injection-recovery tests. We then demonstrate how this tool can be combined with other observational techniques, including high-resolution adaptive optics imaging and time-series radial velocity observations, to comprehensively constrain the probability that a given star hosts an unseen stellar companion. As a proof of concept, we apply this methodology to calculate stellar companion probabilities for an archetypal nearby, Solar-type star.

This paper is organized as follows. In Section 2, we describe our automated SB2-detection algorithm, including spectrum preprocessing, synthetic spectrum generation, peak-finding methodology, and injection-recovery testing. In Section 3, we present a probabilistic framework for combining constraints from multiple observational techniques and apply it to calculate stellar companion probabilities on an example star. We conclude in Section 4 with a summary of our results and a discussion of the implications for future direct imaging missions.

\section{Methodology} \label{sec:method}

In this section, we define our methodology for detecting SB2s in high-resolution spectra. The methodology is largely inspired by previous techniques \citep[e.g.,][]{kolbl2015reamatch}, but is designed with automation in mind. This framework allows for reliable detection and statistically robust rejection of binary star signals.

We tested our algorithm on a spectrum of the Sun collected by SoCal \citep{rubenzahl2023socal}, a solar feed for the Keck Planet Finder (KPF) spectrograph on the 10m Keck-I telescope \citep{gibson2024kpf}. The spectrum was acquired on 2024 June 19 UT and achieved a peak signal-to-noise ratio (S/N) per resolution element of 1300 in the green arm (445–600 nm) and 1900 in the red arm (600–870 nm) after stacking the three science traces. { The spectrum used in this paper has all three traces summed, has outliers (caused by cosmic ray strikes, hot pixels, and insensitive pixels) rejected, and has a wavelength solution defined as the average of the wavelength solutions from the three individual science traces. Further spectrum preprocessing is described below.}

\begin{figure*}[t!]
  \centering
    \includegraphics[width=\textwidth]{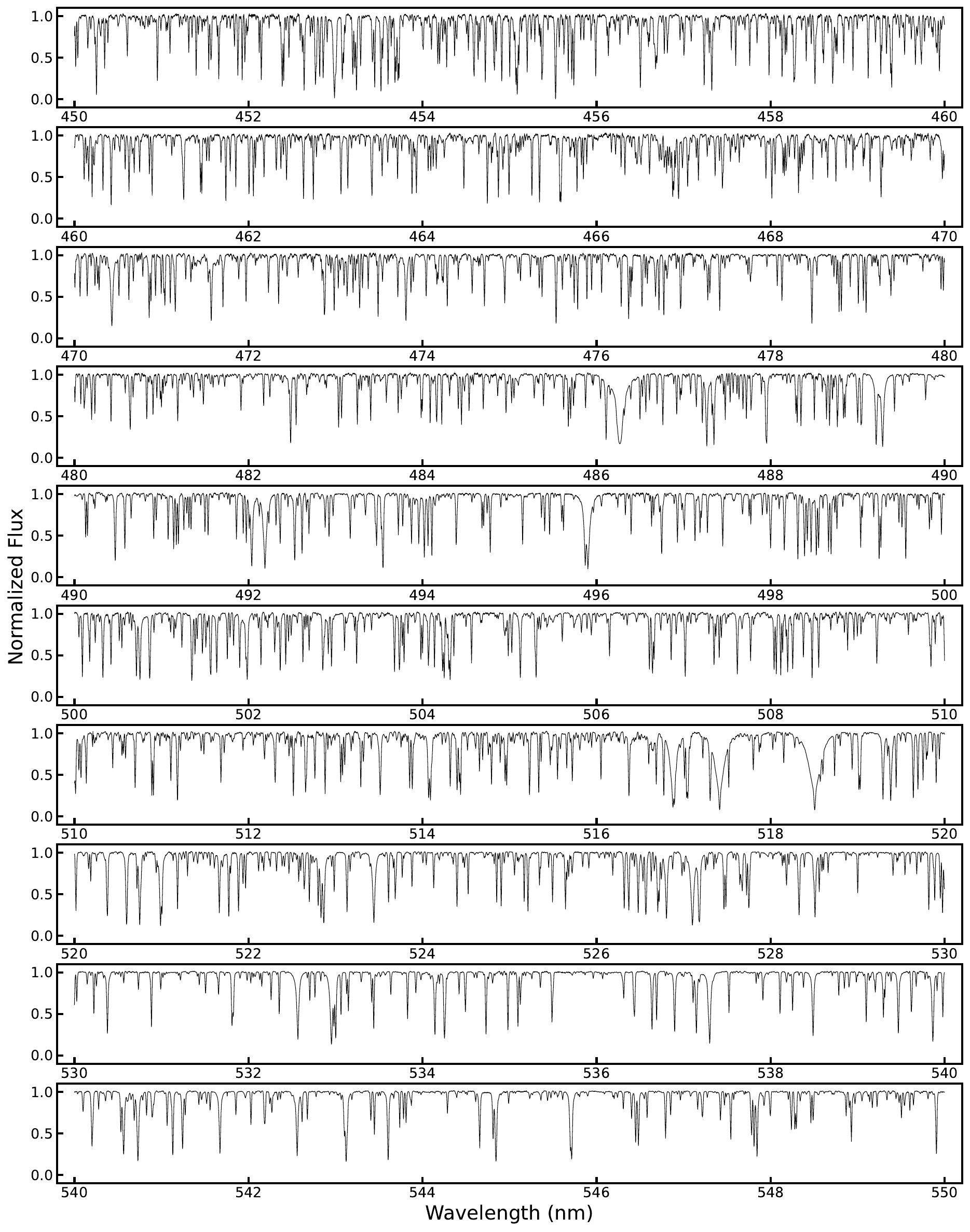}
    \caption{A segment of the stitched and continuum-normalized KPF spectrum of the Sun, which is used throughout this paper for demonstrative purposes. The full spectrum spans 445--870~nm and has had regions of strong telluric absorption masked out.}
    \label{fig:processing}
\end{figure*}

\begin{figure*}[t!]
  \centering
    \includegraphics[width=\textwidth]{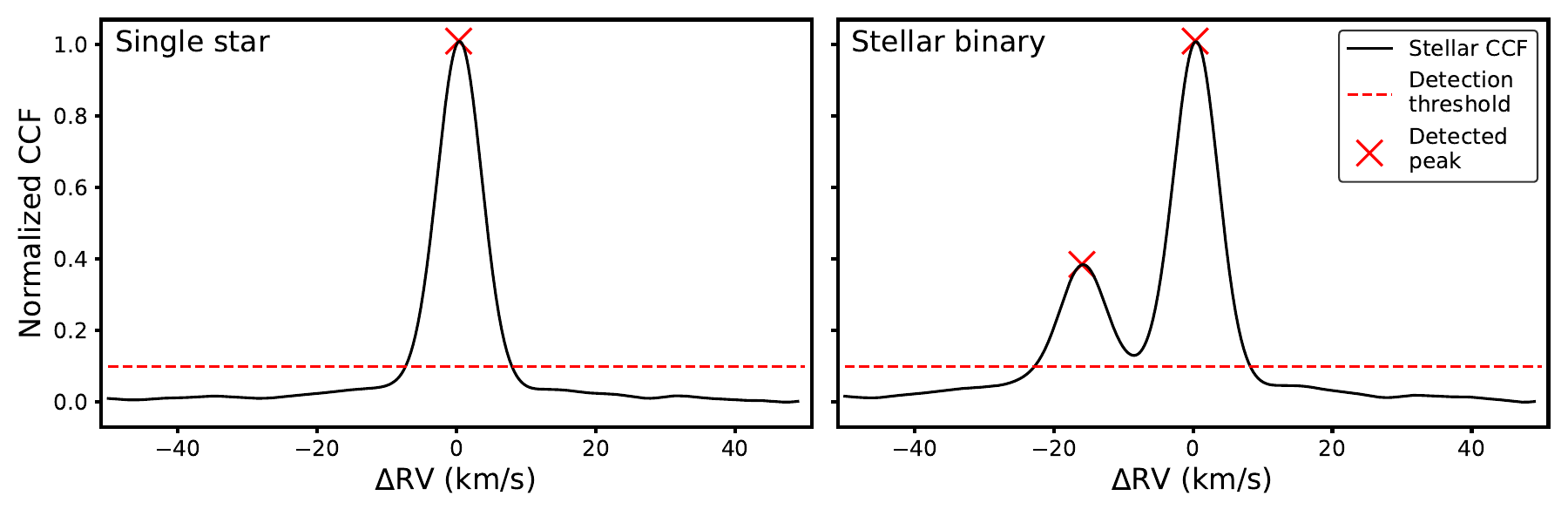}
    \caption{Visualization of the CCF-based SB2-detection technnique employed in this paper. The left-hand panel shows the CCF of a single star { (the Sun)} and the right-hand panel shows the CCF of a pair of spatially unresolved binary stars (the Sun and a synthetic star with $T_{\rm eff} = 5000 \, {\rm K}$, $\log g = 4.5$, and ${\rm [Fe/H] = 0.0}$). The red dashed line is the threshold above which peaks are searched for.}
    \label{fig:ccfs}
\end{figure*}

\subsection{SB2-detection Algorithm}\label{Sec:Algo}

We searched for evidence of SB2s using the CCF of the stellar spectrum. As previous works have demonstrated \citep[e.g.,][]{kolbl2015reamatch}, the presence of a spatially unresolved companion star can be inferred from the CCF based on the number of peaks. A single star contains only a single peak in the CCF, whereas a binary star system contains two peaks, with some offset in radial velocity ($\Delta {\rm RV}$), due to the measured spectrum containing the absorption lines of both stars. 

{ Before calculating CCFs, we performed preprocessing steps on our KPF Solar spectrum. First, we continuum normalized each order using the \texttt{LSQUnivariateSpline} function in the \texttt{interpolate} subpackage of \texttt{SciPy} \citep{virtanen2020SciPy-NMeth}. After flattening each order, we trimmed areas where each order overlapped by removing the redundant data with a lower signal-to-noise ratio. The flattened orders where then stitched together to form a single spectrum spanning a wavelength range of 445--870 nm. Finally, we removed outliers (caused by cosmic ray strikes, hot pixels, and insensitive pixels) by clipping all data points with flux outside the range of [0, 1.1] and masked out regions of the spectrum that were strongly impacted by telluric absorption (e.g., the oxygen A and B bands). A segment of our final processed spectrum is shown in Figure~\ref{fig:processing}. Using our final preprocessed spectrum,} we calculated the CCF by convolving { it} with a binary stellar line mask. In the left-hand panel of Figure~\ref{fig:ccfs}, we show the CCF of our KPF Solar spectrum calculated using a G2V line mask. { The line masks used in this step and the remainder of the analysis are those created by the ESPRESSO pipeline team \citep{pepe2021espresso, figueira2025espresso}, which are optimized on a per-spectral-type basis for precise radial velocity derivation.}

We discerned the number of peaks in any given CCF in an automated way using the $\texttt{signal.find$\_$peaks}$ function in the $\texttt{SciPy}$ Python package \citep{virtanen2020SciPy-NMeth}. In short, this algorithm works by searching for local maxima by comparison of neighboring values. To prevent the algorithm from mistakenly identifying noise in the wings of the CCF as stellar companions (i.e., false alarm detections), we defined a height that the normalized CCF must surpass in order to be considered a peak caused by the presence of additional stellar lines.

The appropriate height threshold is determined on a per-spectrum basis, due to the fact that each individual spectrum will have unique noise properties that can cause false alarm detections. To calculate this threshold, we first generated 100 synthetic spectra with properties identical to that of the target star using \texttt{Starfish} { \citep{czekala2015_starfish, czekala2018_starfish}, which can produce grids of PHOENIX spectra within a defined range of effective temperatures ($T_{\rm eff}$), surface gravities ($\log g$), and metallicities that are convolved to a specified resolving power \citep{husser2013_phoenix}.}  Each synthetic spectrum was then rotationally broadened to the desired projected rotational velocity ($v \sin{i}$) using the technique outlined in \citet{carvalho2023_broaden} and continuum normalized. Next, we injected noise into each synthetic spectrum by resampling each data point from a Gaussian distribution with a variance corresponding to the variance of the observed spectrum at the corresponding wavelength. Lastly, we calculated a unique CCF for each of the synthetic spectra. We ran the peak-finding routine on each of the 100 CCFs, progressively increasing the height threshold of the algorithm from zero until only the central peak is detected in all of the CCFs. This threshold corresponds to that for which the false alarm probability (FAP) falls below $1\%$. 

Once the optimal height threshold was determined, we ran the peak-finding routine on the observed spectrum. In the right-hand panel of Figure~\ref{fig:ccfs}, we show an { illustrative} example of a CCF produced by a { our Solar spectrum and a synthetic binary companion,} { which} is correctly identified as a double-lined spectrum by this technique.

\begin{figure}[t!]
  \centering
    \includegraphics[width=0.49\textwidth]{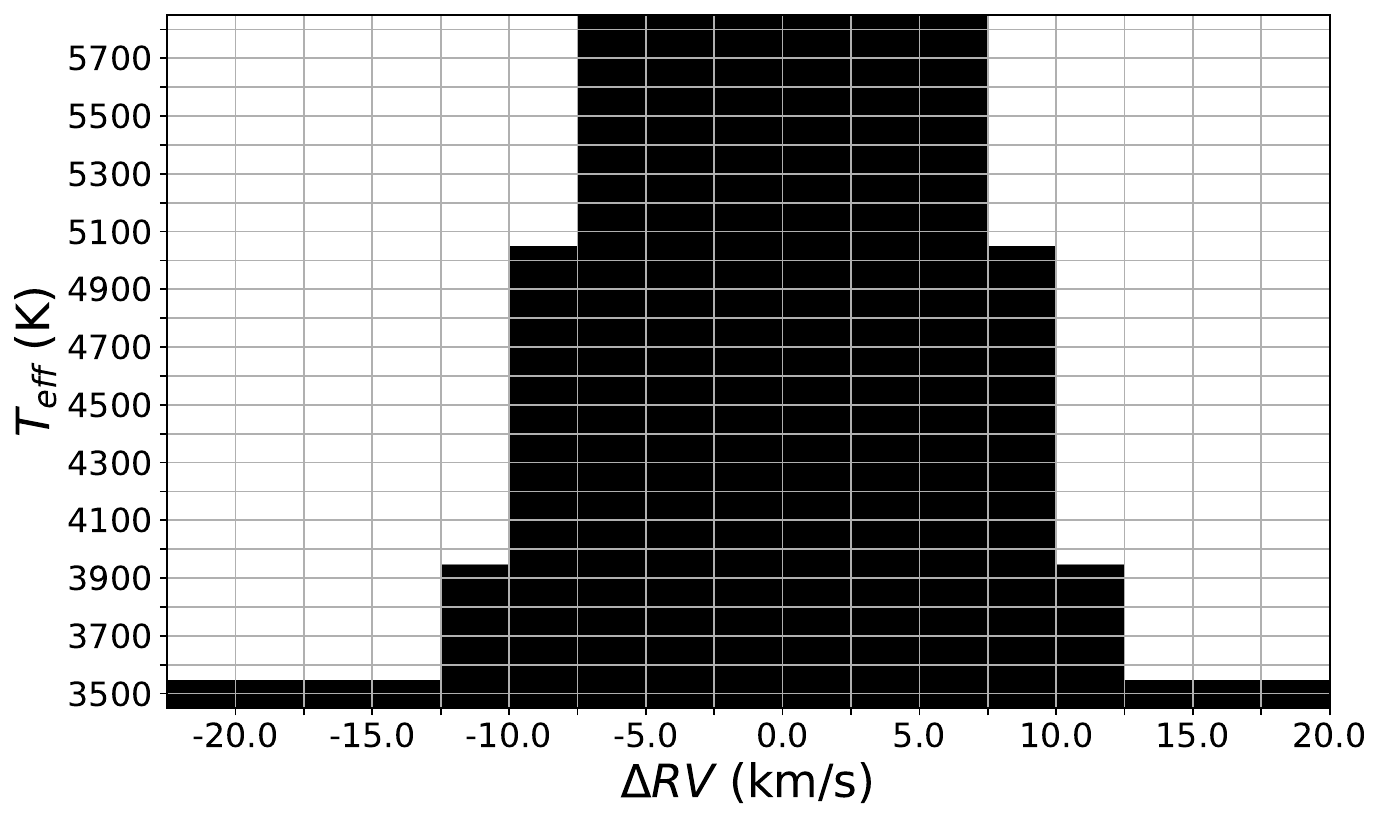}
    \caption{Injection-recovery test results produced using our KPF Solar spectrum and a G2V stellar line mask. Light regions are those in which companion stars are easily detectable and dark regions are those in which companion stars are not. In general, stars with $T_{\rm eff} \gtrsim 3600 \, {\rm K}$ and $|\Delta {\rm RV}| \gtrsim 10 \, {\rm km/s}$ are detectable.}
    \label{fig:injectionrecovery}
\end{figure}

\subsection{Injection-Recovery Tests}\label{sec:injrec}

While the detection of a double-lined spectrum likely indicates the presence of a companion star, the nondetection of a double-lined spectrum does not necessarily mean that the target star is single. Companion stars that have a very small $\Delta {\rm RV}$ or are very faint often produce CCF signals that are either obscured by the central peak of the primary star or exist below the noise level of the CCF wings. In the event of a nondetection, we performed injection-recovery tests on the observed spectrum to determine what types of companion stars we can actually rule out by our technique.

\begin{figure*}[hbtp]
    \centering
    \includegraphics[width=\textwidth]{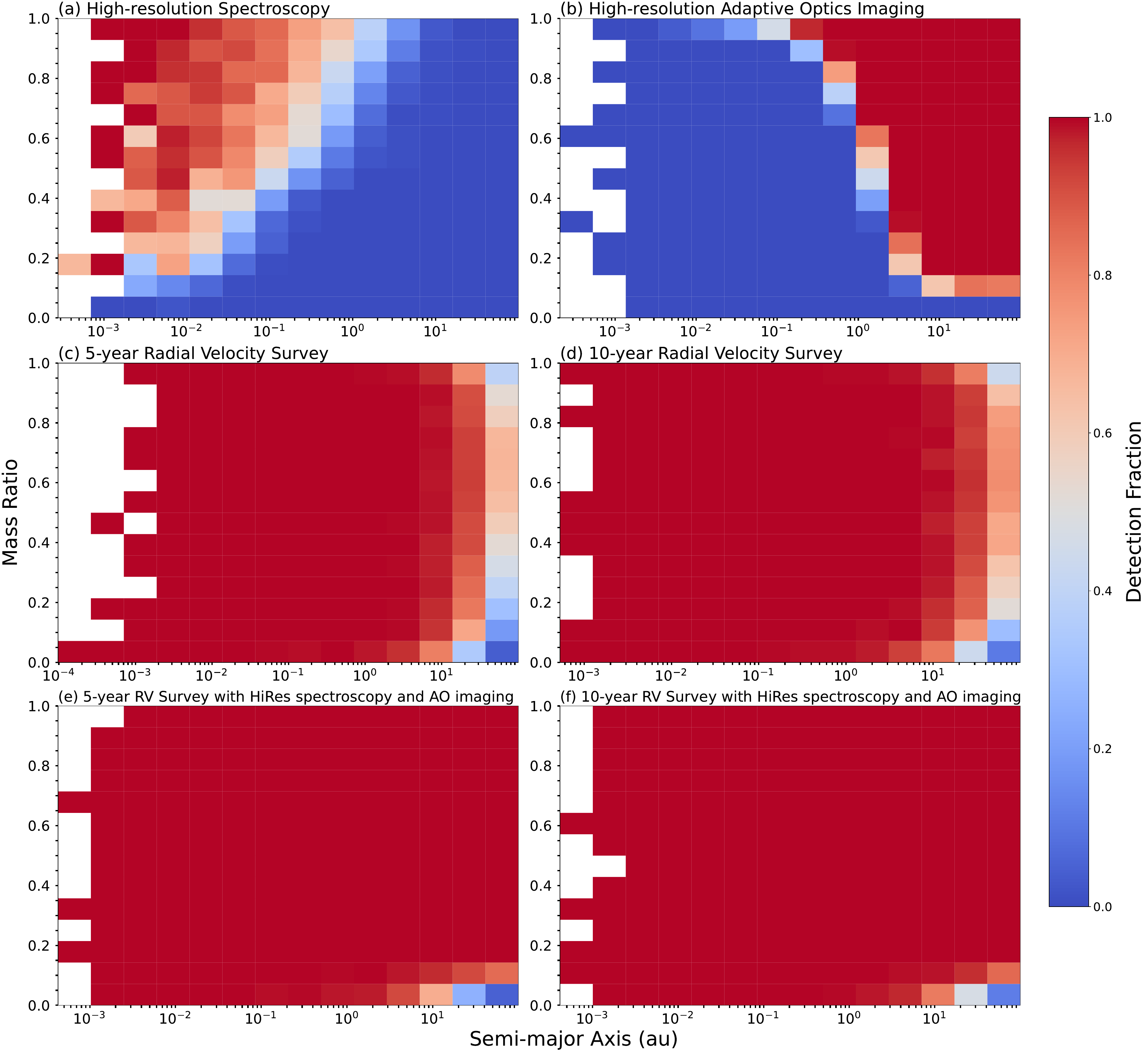}
    \caption{Results of our \texttt{MOLUSC} probabilistic analysis. Each panel represents a different combination of follow-up observations, wherein each cell displays the fraction of detected stellar companions in mass ratio -- semi-major axis space. a) single-epoch high-resolution spectroscopy (46,102/100,000 stellar companions detected), b) high-resolution adaptive optics imaging (79,613/100,000 stellar companions detected), c) 5-year radial velocity survey (92,097/100,000 stellar companions detected), d) 10-year radial velocity survey (94,510/100,000 stellar companions detected), e) 5-year radial velocity survey with single-epoch high-resolution spectroscopy and high-resolution adaptive optics imaging (98,374/100,000 stellar companions detected), f) 10-year radial velocity survey with single-epoch high-resolution spectroscopy and high-resolution adaptive optics imaging (98,720/100,000 stellar companions detected). When all techniques are combined, nearly all undetected stellar companions are M dwarfs with mass ratios of $< 0.1$ and semimajor axes between $10$ and $100$~au.}\label{fig:scatter}
\end{figure*}

We began by generating a grid of synthetic spectra using \texttt{Starfish} across an array of $T_{\rm eff}$, $\log g$, and $\Delta {\rm RV}$. { We held the metallicity of each spectrum equal to that of the primary star (in this case, the Sun). To allow for variations in $\log g$ for stars of different spectra types, we adopted a simple piecewise function where stars with $T_{\rm eff} \leq 3900$~K have $\log g = 5.0$ and stars with $T_{\rm eff} > 3900$~K have $\log g = 4.5$. To account for rotation, we applied a broadening kernel corresponding to $v \sin{i} = 2 \, {\rm km} \, {\rm s}^{-1}$ for all stellar companions. We assumed that the inclinations of the spin axes of the two components are equal, which is supported by recent observations that most close stellar binaries are spin-orbit aligned \citep{marcussen2022spinorbit, smith2024spinorbit}.\footnote{This assumption is not valid for binary companions with much wider ($a > 10$~au) orbits, but our technique is not sensitive to those widely separated companions anyway (see Figure~\ref{fig:scatter}).} We note that Solar-age stars cooler than the Sun should actually have lower values of $v \sin{i}$ due to smaller radii and longer rotation periods, but these values are too small to be resolved in our spectra (i.e., instrumental broadening dominates), so a more realistic treatment of stellar rotation would not influence our results. In reality, both $\log g$ and $v \sin{i}$ are dependent on age; young stars will have higher values of both $\log g$ and $v \sin{i}$ and would therefore require a more careful treatment of these parameters. That said, we stress that our analysis is only valid for stars with ages similar to the Sun and older. Next, each synthetic spectrum was convolved to the resolution of KPF and continuum normalized using the spine-fitting technique described above. Lastly, prior to injecting the synthetic companion into the observed spectrum with some $\Delta {\rm RV}$, we corrected for wavelength-dependent flux ratio between the two stars, which is simply calculated as the ratio of the best-fit continuum spline of the companion to the best-fit continuum spline of a synthetic Sun-like star, where each synthetic spectrum is multiplied by the squared radius of the star (in order to convert from surface flux to flux density).}

We ran the peak-finding routine on each of the composite spectra, producing the detection maps shown in Figure~\ref{fig:injectionrecovery}. In these figures, light regions correspond to companion stars with combinations of $T_{\rm eff}$ and $\Delta {\rm RV}$ that are detectable, whereas dark regions correspond to companion stars with combinations of $T_{\rm eff}$ and $\Delta {\rm RV}$ that are not detectable. Using our KPF Solar spectrum, we found that companion stars are only detectable if they have $T_{\rm eff} \gtrsim 3600 \, {\rm K}$ and $|\Delta {\rm RV}| \gtrsim 10 \, {\rm km/s}$. We performed this calculation using a G2V stellar line mask.

\section{Probabilistic Analysis}

By construction, our automated SB2-detection algorithm is reliable (${\rm FAP} < 1 \%$) and has a well-understood completeness. This means that the tool can be used to calculate the probability that any given star has a binary star based on the nondetection of an SB2, assuming the prior distribution of companion stars in $T_{\rm eff}$ and $\Delta {\rm RV}$ space is known. We can perform this experiment using \texttt{MOLUSC} \citep{wood2021molusc}, a Python-based code that simulates a realistic population of stellar companions based on the properties (e.g., age and mass) of the primary star. { These populations of stellar companions are generated using prior distributions on orbital period (a lognormal distribution similar to that in \citealt{raghavan2010multiplicity}), orbital eccentricity (a period-dependent normal distribution, where the mean the standard deviation of the distribution are proportional to $\log_{10} P$), and mass ratio (a uniform distribution, based on the findings of \citealt{moe2017pq, elbadry2019twin, kounkel2019}). Prior distributions for remaining orbital parameters (e.g., orbital inclination) are based on standard geometrical arguments.} \texttt{MOLUSC} also has the ability to rule out a fraction of stellar companions based on user-supplied constraints, such as limits obtained from high-resolution adaptive optics imaging and time-series radial velocity observations. In this section, we utilized \texttt{MOLUSC} in tandem with our SB2-detection algorithm to calculate the probability that a given Sun-like star has an unseen stellar companion, aiming to achieve a stellar companion probability of $<1 \%$.

We initialized \texttt{MOLUSC} using a $1 \, M_{\odot}$ star with an age of 5~Gyr and the coordinates of HD~109085. We generated $10^5$ companions and determined what fraction of companions can be eliminated based on different forms of follow-up observations. We first eliminated all companions with projected separations greater than 100~au, which corresponds to $2''$ for a star at a distance of 50~pc. We then considered the following observations: single-epoch high-resolution spectroscopy (i.e., a single spectrum), high-resolution adaptive optics imaging, and long-baseline time-series radial velocities. After running \texttt{MOLUSC} for each case, we calculated the probability of there being an undetected stellar companion companion by multiplying the fraction of surviving companions by 0.5 (corresponding approximately to the stellar multiplicity rate for G-type stars; \citealt{raghavan2010multiplicity, gonzalezpayo2026}). Our results are summarized in Table~\ref{tab1}, which reports the combination of observations used, the number of surviving stellar companions (out of $10^5$), and the remaining stellar companion probability.

\subsection{Single-Epoch High-Resolution Spectroscopy}

We first determined the stellar companion probability using a hypothetical high-resolution, high-S/N spectrum with visible wavelength coverage (i.e., the KPF Solar spectrum discussed above). For each companion star returned by \texttt{MOLUSC}, we calculated $\Delta {\rm RV}$ and estimated $T_{\rm eff}$ based on the secondary-primary mass ratio using the relations in \citet{pecaut2013table}.\footnote{\url{https://www.pas.rochester.edu/~emamajek/EEM_dwarf_UBVIJHK_colors_Teff.txt}} We then compared these values with the injection-recovery map in Figure~\ref{fig:injectionrecovery} to determine whether each star is detectable. This process ruled our $46.1 \%$ of stellar companions around the star. The locations of the ruled out companions in mass -- orbital separation space are shown in the top-left panel of Figure~\ref{fig:scatter}. In general, this method is capable of efficiently ruling out stellar companions out to 1~au for mass ratios down to $0.8$ and companions out to 0.1~au for mass ratios down to $0.4$.

\subsection{High-Resolution Adaptive Optics Imaging}

Next, we explored the role of high-resolution adaptive optics imaging plays for ruling out stellar companions in these systems. To do this, we created a contrast curve extracted from a hypothetical $K$-band image of the star, which achieves a contrast of 6.25 at 0.5 arcseconds, 6.8 at 1 arcsecond, and 7.0 beyond 1.4 arcseconds. { This contrast curve is intended to resemble those taken by instruments like Keck/NIRC2 \citep[e.g.,][]{schlieder2021nirc2}.} We input this contrast curve directly into \texttt{MOLUSC}, which has a built-in algorithm for ruling out stellar companions based on these follow-up observations. This process ruled our $79.6 \%$ of stellar companions around the star. The locations of the ruled out companions in mass -- orbital separation space are shown in the top-right panel of Figure~\ref{fig:scatter}. This method excels at ruling out stellar companions beyond 5~au for mass ratios down to $0.1$, but misses stellar companions that are close to conjunction.

\input{Table1}

\subsection{Time-Series Radial Velocities}

Long-baseline radial velocity surveys are likely to plan an important role in HWO target selection, due to their abilities to detect relatively low-mass companions on wide orbits \citep[e.g.,][]{rosenthal2021CLS, harada2024SPORESII}. The presence of giant planets or low-mass stellar companions near the habitable zone can render earth-like planet dynamically unstable, making their identification crucial for optimizing the scientific output of the mission \citep[e.g.,][]{kane2024perturbers}.

To investigate the sensitivity of radial velocity surveys to low-mass stellar companions, we reran MOLUSC using two hypothetical radial velocity data sets: (1) a 5-year survey in which the star is observed twice per month over repeated 6-month periods, and (2) a 10-year survey in which the star is observed twice per month over repeated 6-month periods. The timings of the two observations in any given month are chosen at random to avoid introducing aliases that would hide companions with specific orbital periods. The data is assumed to be invariant with time (i.e., there is no companion) with white noise drawn from a Gaussian distribution with a standard deviation of 2~m~s$^{-1}$ and an added jitter of 1~m~s$^{-1}$. We assume that this uncertainty encapsulates both instrumental and stellar variability. The spectrograph is assumed to have a resolution of 100,000.

The results of these two analyses are shown in Figure~\ref{fig:scatter}. The 5-year survey is capable of efficiently ruling out stellar-mass companions with mass ratios down to 0.1 and orbital separations out to $5$~au. The 10-year survey is capable of efficiently ruling out stellar-mass companions with masses down to 0.1 and orbital separations out to $10$~au. The two surveys ruled out $92.1\%$ and $94.5\%$ of stellar companions, respectively (see Table~\ref{tab1}).

\begin{figure*}[t!]
  \centering
    \includegraphics[width=\textwidth]{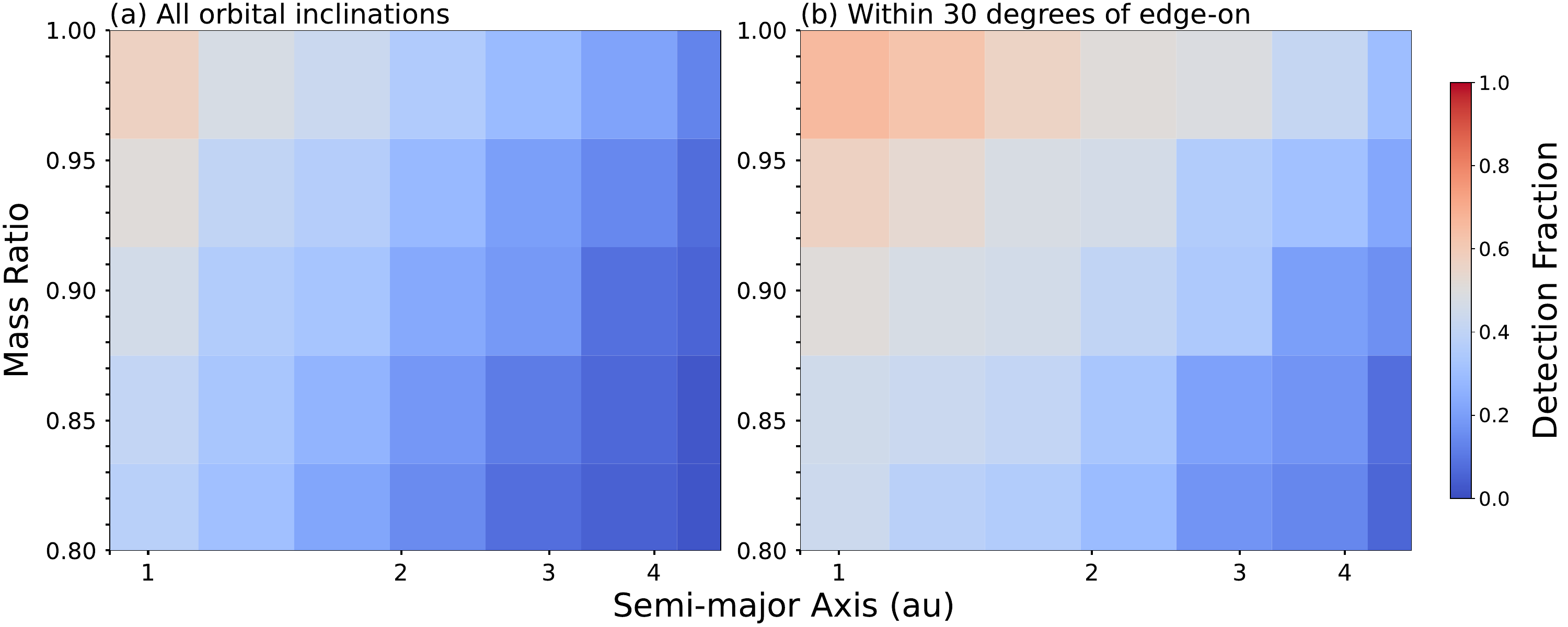}
    \caption{Panel (a) of Figure~\ref{fig:scatter}, zoomed in to the region of parameter space that will contain the majority of planet false positives in Gaia DR4. Our SB2-detection algorithm is capable of detecting up to $50\%$ of false positives when all orbital inclinations are allowed and up to $70\%$ of false positives when inclinations are restricted to within $30^\circ$ of edge-on.}\label{fig:gaia}
\end{figure*}

\subsection{Combined Results}

Based on this analysis, we see that different forms of follow-up observations excel at rejecting stellar companions in different mass and orbital separation regimes. By combining all three methods, we can obtain the most stringent constraints on stellar companion probability. In Table~\ref{tab1}, we report these combined probabilities for the two different radial velocity data sets. In short, the 5-year and 10-year radial velocity surveys, when combined with single-epoch high-resolution spectroscopy and high-resolution adaptive optics imaging, ruled out $98.4\%$ and $98.7\%$ of stellar companions, respectively. { After multiplying the fraction of remaining companions by our prior of 0.5, we obtain stellar companion probabilities of $<1\%$ for both scenarios.} Surviving stellar companions tend to be low-mass stars with mass ratios $<0.1$ with semimajor axes between 10 and 100~au (see Figure~\ref{fig:scatter}). We note that it may be possible to rule out these low-mass companions using techniques not explored in this paper, such as by searching for astrometric accelerations using \textit{Gaia}-\textit{Hipparcos} data \citep[e.g.,][]{2018ApJS..239...31B, 2021ApJS..254...42B, 2020AJ....160..240X, 2021AJ....162...12V, 2022A&A...657A...7K, 2023ApJ...950L..19F, 2023MNRAS.522.5622L, 2024MNRAS.533.3501L}, but those methods are beyond the scope of this paper.

\subsection{ Applications to Gaia DR4 Planet Candidate Vetting}

{ Between $20\%$ and $50\%$ of astrometric planet candidates in Gaia DR4 are expected to be astrophysical false positives (depending on host spectral type; \citealt{lammers2026gaia}), with the most common false positive scenario being that of a roughly equal-mass binary that is spatially unresolved \citep{marcussen2023gaia}. High-resolution spectroscopic survey will likely play an important role in identifying these false positives, which can appear as SB2s. To give an idea of the sensitivity of our algorithm to these systems, Figure~\ref{fig:gaia} shows a cropped version of panel (a) of Figure~\ref{fig:scatter} zoomed in to the region of parameter space in which most of these false positives will reside. The left-hand panel of Figure~\ref{fig:gaia} shows that for Sun-like stars, between $20\%$ and $50\%$ of these binaries can be ruled out with a single high-S/N, high-resolution KPF spectrum. The right-hand panel performs the same calculation, but restricts the population of simulated systems to those with inclinations within $30^\circ$ of edge-on, demonstrating that we achieve a boost in detection efficiency ($40\% - 70\%$) due to the removal of edge-on systems with relatively low $\Delta$RV values. Because edge-on planetary systems will be among the best targets for radial velocity observations capable of confirming planets and refining their orbits, this technique will be a useful tool for prioritizing follow-up targets.}

\section{Discussion and Conclusions}

We have introduced an automated SB2-detection algorithm using a CCF peak-finding technique. Our algorithm calculates the completeness of SB2 detection using injection-recovery tests, enabling robust statistical analyses of binary star probability. We tested our algorithm on a hypothetical HWO target star using MOLUSC, a code that simulates realistic binary populations given stellar parameters like mass and age, and found that it can rule out $46.1\%$ of stellar-mass companions. However, the technique is only efficient at detecting companions out to 1~au for mass ratios down to $0.8$ and companions out to 0.1~au for mass ratios down to $0.4$. To determine how this technique could be combined with other observation to obtain more stringent constraints on binary probability, we also folded in high-resolution adaptive optics imaging data and long-baseline radial velocity data. We found that the high-resolution adaptive optics imaging data can rule out $79.6\%$ of stellar companions around the typical nearby, Sun-like star, excelling at ruling out companion stars with mass ratio down to 0.1 and orbital separation beyond 5~au. The radial velocity surveys performed the best of the three observations, { ruling out  $92.1\%$ and $94.5\%$ for 5-year and 10-year surveys}, respectively, with a high sensitivity to stellar-mass companions of all masses. When combined, three three techniques achieve a stellar companion probability of $< 1\%$ for nearby stars.

This study highlights the importance of diverse and long-baseline reconnaissance observations for target selection for future missions, such as HWO. When combining several different techniques, we are able to determine which stars have the lowest probabilities of hosting unknown stellar companions, thereby determining the best stars for more careful and curated follow-up observations.

\begin{acknowledgements}
S.G.\ is supported by an NSF Astronomy and Astrophysics Postdoctoral Fellowship under award AST-2303922. A.W.H.\ acknowledges funding support from NASA grant No.\ 80NSSC24K0161.

Some of the data presented herein were obtained at the W.\,M.~Keck Observatory, which is operated as a scientific partnership among the California Institute of Technology, the University of California, and the National Aeronautics and Space Administration. We wish to recognize and acknowledge the very significant cultural role and reverence that the summit of Maunakea has always had within the indigenous Hawaiian community. We are most fortunate to have the opportunity to conduct observations from this mountain.
\end{acknowledgements}

\facilities{Keck I (KPF)}

\software{MOLUSC \citep{wood2021molusc}, \texttt{SciPy} \citep{virtanen2020SciPy}}

\newpage

\bibliography{bibliography}{}
\bibliographystyle{aasjournal}

\end{document}

%% file: Table1.tex
\begin{deluxetable*}{lcc}[t!] \label{tab1}
\centering
\tabletypesize{\scriptsize}
\tablewidth{0pt}
\tablecaption{Summary of \texttt{MOLUSC} Probabilistic Analysis} 
\tablehead{\colhead{Follow-up Observations Included} & \colhead{Number of Surviving Companions} & \colhead{Stellar Companion Probability} }
\startdata
Hi-Res Spectroscopy & 53,898 & 0.269 \\
AO Imaging & 20,387 & 0.102 \\
5-year RV Survey & 7,903 & 0.040 \\
10-year RV Survey & 5,490 & 0.027 \\
5-year RV Survey + Hi-Res Spectroscopy + AO Imaging & 1,626 & 0.008 \\
10-year RV Survey + Hi-Res Spectroscopy + AO Imaging & 1,280 & 0.006 \\
\enddata
\end{deluxetable*}